\documentclass[superscriptaddress,secnumarabic,nofootinbib,onecolumn]{revtex4}
\usepackage[english]{babel}
\usepackage{amsmath,amssymb} 
\usepackage{stmaryrd,dsfont,txfonts}
\usepackage{subfigure}
\usepackage{graphicx}
\usepackage{dcolumn}
\usepackage[sort&compress]{natbib}
\usepackage{float}
\usepackage{ifpdf}
\ifpdf
  \usepackage[dvipdfm,colorlinks,hyperindex]{hyperref}
\else
  \usepackage[hypertex,colorlinks,hyperindex]{hyperref}
\fi

\DeclareMathAlphabet{\mathvec}{OT1}{cmr}{bx}{sl}
\SetMathAlphabet{\mathvec}{bold}{OT1}{cmr}{bx}{sl}
\renewcommand{\vec}[1]{\mathvec{#1}}

\begin{document}

\title{Spin orbit interaction and zitterbewegung in symmetric wells}
\author{Esmerindo Bernardes}
\email{sousa@if.sc.usp.br}
\affiliation{Instituto de F{\'{\i}}sica de S{\~{a}}o Carlos\\
  Universidade de S{\~{a}}o Paulo\\
  Av. do Trabalhador S{\~{a}}o-carlense, 400 CP 369\\
  13560.970 S{\~{a}}o Carlos, SP, Brazil}
\author{John Schliemann}
\affiliation{Institute for Theoretical Physics\\
  University of Regensburg\\
  D-93040 Regensburg, Germany}
\affiliation{Department of Physics and Astronomy\\
  University of Basel\\
  CH-4056 Basel, Switzerland}
\author{J. Carlos Egues}
\email{egues@if.sc.usp.br}
\affiliation{Instituto de F{\'{\i}}sica de S{\~{a}}o Carlos\\
  Universidade de S{\~{a}}o Paulo\\
  Av. do Trabalhador S{\~{a}}o-carlense, 400 CP 369\\
  13560.970 S{\~{a}}o Carlos, SP, Brazil}
\affiliation{Department of Physics and Astronomy\\
  University of Basel\\
  CH-4056 Basel, SwitzerlandK}
\affiliation{Kavli Institute for Theoretical Physics\\
  University of California, Santa Barbara,  93106 California, USA}
\author{Daniel Loss}
\affiliation{Department of Physics and Astronomy\\
  University of Basel\\
  CH-4056 Basel, Switzerland}
\affiliation{Kavli Institute for Theoretical Physics\\
  University of California, Santa Barbara,  93106 California, USA}
\date{\today}

\date{\today }

\begin{abstract}
  Recently, we have introduced a novel inter-subband-induced
  spin-orbit (s-o) coupling [Phys. Rev. Lett. 99, 076603 (2007);
  cond-mat/0607218] arising in \textit{symmetric} wells with at least
  two subbands. This new s-o coupling gives rise to an usual
  zitterbewegung -- i.e. the semiconductor analog to the relativistic
  trembling motion of electrons -- with cycloidal motion without
  magnetic fields. Here we complement these findings by explicitly
  deriving expressions for the corresponding zitterbewegung in spin
  space.
\end{abstract}

\maketitle


\section{Introduction}

A controllable coupling between the spin and the orbital degrees of
freedom in nanostructures is a highly desirable ingredient in the
emerging fields of semiconductor spintronics and spin-based quantum
computation and communication \cite{overview}. The gate-tunable
\cite{nitta} Rashba s-o interaction \cite{Rashba60} present in
two-dimensional electron gases (2DEGs) formed in structurally
asymmetric confining potentials offers such a possibility. The
seminal proposal of Datta and Das \cite{datta} of a spin field
effect transistor highlights the use of the Rashba spin-orbit
interaction to coherently rotate spins. More recently, the Rashba
s-o interaction has been proposed as a convenient coherent mechanism
for manipulating entangled electrons (``flying qubits'') in the
solid state \cite{ebl}. Interestingly, systems with bulk inversion
asymmetry exhibit an additional spin orbit interaction, termed the
Dresselhaus s-o \cite{Dresselhaus55}, whose interplay with the
Rashba s-o can be used to devise a robust spin field effect
transistor \cite{Schliemann03a}.

So far, s-o effects in 2DEGs have only been investigated in
(asymmetric) quantum-well systems with a single confined state.
Recently, we have introduced a new type of spin-orbit interaction
present in quantum wells with two confined subbands
\cite{Bernardes}. Unlike the ordinary Rashba s-o interaction, ours
is non-zero even in symmetric wells -- as it arises from the
inter-subband-coupling between the lowest (even) and the first (odd)
excited states of the well. As pointed out in Ref. \cite{Bernardes},
this new \textit{inter-subband-induced} s-o coupling leads to an
unusual dynamics of injected spin-polarized wave packets -- the
zitterbewegung or trembling motion -- with cycloidal trajectories
without magnetic fields. This is qualitatively different from the
zitterbewegung in the presence of Rashba or Dresselhaus s-o
interactions \cite{zitter1}. Here we complement the investigation in
Ref. \cite{Bernardes} by deriving explicit expressions for the
zittebewegung in spin space, i.e., the spin dynamics of the injected
electron. In what follows, we first present our new Hamiltonian with
its eigenvalues and the corresponding time-evolution operator, and
then proceed to determine the spin dynamics of injected electrons.

\section{Effective Hamiltonian: novel spin-orbit term}

Starting from the usual $8\times 8$ Kane model one can derive a
$2\times 2$ effective Schr\"{o}dinger equation for the conduction
electrons by a folding down process. We have performed this
procedure \cite{Bernardes} for a symmetric quantum well with two
subbands with edges at the quantized energies $\varepsilon_e$
(lowest level) and $\varepsilon_o$ (first level). In terms of the
real-spin $\sigma$ and pseudo-spin $\tau$ (describing the subband
degree of freedom) Pauli matrices, our $4\times 4$ Hamiltonian for
an electrons with effective mass $m^*$ can be cast in the compact
form
\begin{equation}
\mathcal{H} = \left( \frac{p_{\shortparallel}^{2}}{2m^{*}}
+\epsilon_{+} \right) \mathbf{1}\otimes\mathbf{1} - \epsilon_{-}
\tau^{z}\otimes\mathbf{1} +\frac{\eta}{\hbar}\tau^{x}\otimes
\left(p_{x}\sigma^{y}-p_{y}\sigma^{x}\right),\quad
\epsilon_{\pm}=\frac{\varepsilon_{o}\pm\varepsilon_{e}}{2},
\label{eq1}
\end{equation}
where the new inter-subband-induced s-o coupling $\eta$ is given by
\begin{equation}
\eta =-\left( \frac{1}{E_{g}^{2}}-\frac{1}{\left( E_{g}+\Delta \right) ^{2}}%
\right) \frac{P^{2}}{3}\langle e|\partial _{z}V(z)|o\rangle \, +\left( \frac{%
\delta _{V}}{E_{g}^{2}}-\frac{\delta _{\Delta }}{\left( E_{g}+\Delta \right)
^{2}}\right) \frac{P^{2}}{3}\langle e|\partial _{z}h(z)|o\rangle,
\label{eta}
\end{equation}
where the Kane matrix element is $P=-i\hbar \langle S|p_{x}|X\rangle /m_{0}$%
, with $|S\rangle $ and $|X\rangle $ being the band-edge ($\Gamma $ point)
periodic Bloch functions, the fundamental and split-off band gaps in the
well region are $E_g$ and $\Delta$, respectively, $|e\rangle$ and $|o\rangle
$ denote the confined ground and first excited states of the well, $V(z)$ is
the Hartree-type contribution to the potential, $h(z)$ is the profile
function defining the quantum well, and $\delta_v$, $\delta_\Delta$ are the
offsets between the valence bands in the well and barrier regions.

Note that $\mathcal{H}$ bears a close similarity to the usual Rashba
Hamiltonian \cite{Rashba60}; however, the corresponding s-o coupling $\eta$
can be non-zero even for symmetric wells -- this can be easily understood by
noting that $\eta$ contains matrix elements between the ground and excited
states. For a symmetric square-well potential, for instance, the first term
in Eq. (\ref{eta}) vanishes (i.e. there is no Hartree contribution) and the
second term reduces to
\begin{equation}
\eta =\frac{2P^{2}}{3}\left( \frac{\delta _{V}}{E_{g}^{2}}-\frac{\delta
_{\Delta }}{\left( E_{g}+\Delta \right) ^{2}}\right)
\varphi_o(a)\varphi_e(a),
\end{equation}
where $\varphi_i(z)$ $i=e,o$ denote the confined well wave functions. Note
that $\partial _{z}h(z)=$ $-\delta (z-a)+\delta (z-a)$ since $%
h(z)=\Theta(a-z) + \Theta(z-a)$ [$\Theta(z)$: Heaviside function] defines a
square well of width $L=2a$. As shown in Ref. (\cite{Bernardes}) the
magnitude of $\eta$ can be comparable to that of the usual Rashba s-o
coupling.

The Hamiltonian \eqref{eq1} can be easily diagonalized yielding the
eigenvalues
\begin{equation}  \label{eq:E}
\mathcal{E}_{\pm} = \varepsilon_k \pm\hbar\Omega,
\end{equation}
where $\varepsilon_k=\frac{\hbar^2 k_{\shortparallel}^{2}}{2m^{*}} +
\epsilon_{+}$ and $\hbar\Omega=\sqrt{\epsilon_{-}^{2}+\eta^2k_{%
\shortparallel}^{2}}$.

\section{Time evolution operator}

The time evolution operator $U=\exp (-i\mathcal{H}t/\hbar )$ can be
straightforwardly obtained from Eq. (\ref{eq1}). In the plane wave basis $%
\{|k_{\shortparallel },\sigma _{z}\rangle \}$, we find
\begin{equation}
U=e^{-i\varepsilon _{k}t/\hbar }\left\{ \cos (\Omega t)\mathbf{1}\otimes
\mathbf{1}+i\left[ \epsilon _{-}\tau ^{z}\otimes \mathbf{1}-\eta \tau
^{x}\otimes (k_{x}\sigma ^{y}-k_{y}\sigma ^{x})\right] \frac{\sin (\Omega t)%
}{\hbar \Omega }\right\} .  \label{time-evol}
\end{equation}%
Next we use $U$ to calculate the time evolution of the spin operators.

\section{Zitterbewegung in spin space}

To determine the spin dynamics in our system we have to calculate the time
evolution of the tensor product $\mathbf{1}\otimes \sigma ^{i}$, $i=x,y,z$.
In the Heisenberg picture we have $\sigma _{H}^{i}(t)=U^{\dag}\mathbf{1}%
\otimes \sigma ^{i}U$, where $\sigma ^{i}$'s denote the Pauli matrices at $%
t=0$. For the $z$ component we find
\begin{eqnarray}
\sigma _{H}^{z}(t) &=&\left\{ \cos (\Omega t)\mathbf{1}\otimes \mathbf{1}-i%
\left[ \epsilon _{-}\tau ^{z}\otimes \mathbf{1}-\eta \tau ^{x}\otimes
(k_{x}\sigma ^{y}-k_{y}\sigma ^{x})\right] \frac{\sin (\Omega t)}{\hbar
\Omega }\right\} \times  \notag \\
&&\mathbf{1}\otimes \sigma ^{z}\left\{ \cos (\Omega
t)\mathbf{1}\otimes \mathbf{1}+i\left[ \epsilon _{-}\tau ^{z}\otimes
\mathbf{1}-\eta \tau
^{x}\otimes (k_{x}\sigma ^{y}-k_{y}\sigma ^{x})\right] \frac{\sin (\Omega t)%
}{\hbar \Omega }\right\}  \label{sigmaz1}
\end{eqnarray}
\begin{equation}
\sigma _{H}^{z}(t)=\mathbf{1}\otimes \sigma ^{z}-\frac{\eta }{\hbar \Omega }%
\tau ^{x}\otimes (k_{x}\sigma ^{x}+k_{y}\sigma ^{x})\sin (2\Omega t)-\frac{2%
}{\hbar ^{2}\Omega ^{2}}\left[ \epsilon _{-}\eta \tau ^{y}\otimes
(k_{x}\sigma ^{x}+k_{y}\sigma ^{x})+\eta ^{2}k^{2}\right] \sin ^{2}(\Omega
t).  \label{sigmaz}
\end{equation}
Similarly, for the $x$ and $y$ components we obtain
\begin{equation}
\sigma _{H}^{x}(t)=\mathbf{1}\otimes \sigma ^{x}+\frac{\eta k_{x}}{\hbar
\Omega }\tau ^{x}\otimes \sigma ^{z}\sin (2\Omega t)+\frac{\eta k_{x}}{\hbar
^{2}\Omega ^{2}}\left[ 2\epsilon _{-}\tau ^{y}\otimes \sigma ^{z}-\eta
\mathbf{1}\otimes (k_{x}\sigma ^{x}+k_{y}\sigma ^{x})\right] \sin
^{2}(\Omega t),  \label{sigmax}
\end{equation}%
\begin{equation}
\sigma _{H}^{y}(t)=\mathbf{1}\otimes \sigma ^{y}+\frac{\eta k_{y}}{\hbar
\Omega }\tau ^{x}\otimes \sigma ^{z}\sin (2\Omega t)+\frac{\eta k_{y}}{\hbar
^{2}\Omega ^{2}}\left[ 2\epsilon _{-}\tau ^{y}\otimes \sigma ^{z}-\eta
\mathbf{1}\otimes (k_{x}\sigma ^{x}+k_{y}\sigma ^{x})\right] \sin
^{2}(\Omega t).  \label{sigmay}
\end{equation}
In deriving the above equations we have used the algebraic relations
$(\sigma^i)^2=\mathbf{1}$, $i=x,y,z$ and $\sigma^i
\sigma^j=-\sigma^j\sigma^i =i\sigma^k$, $i,j,k=x,y,z$ (or any cyclic
permutation), obeyed by the Pauli matrices. The $\tau^i$, $i=x,y,z$,
also obey these relations. Equations (\ref{sigmaz})--(\ref{sigmay})
describe the spin dynamics of injected electrons due to the s-o
coupling $\eta $. Together with the equations for $x_{H}(t)$ and
$y_{H}(t)$ derived in Ref. \cite{Bernardes}, we have now the
complete dynamics of an injected electron -- i.e., the time
dependence of both the spin and orbital degrees of freedom. As we
discuss next, this dynamics are intrinsically linked due to the
spin-orbit coupling and the fact that the $z$ component of the total
angular momentum $\vec j$ is conserved.

\section{Spin polarized injection and discussion}

Similarly to Ref. \cite{Bernardes} let us consider the case of a
wide wave packet (approximated by a plane wave) initially injected
into the lowest subband with spin up and an initial (group) velocity
$\vec{v}_{g}= \hat{y}\hbar k_{0y}/m^*$ along the $y$ axis.
In this case we find from the above equations%
\begin{equation}
\langle \sigma _{H}^{z}(t)\rangle =1-\frac{2\eta ^{2}k_{0y}^{2}}{\hbar
^{2}\Omega ^{2}}\sin ^{2}(\Omega t),  \label{exp-sigz}
\end{equation}%
\begin{equation}
\langle \sigma _{H}^{x}(t)\rangle =\langle \sigma _{H}^{y}(t)\rangle =0.
\label{sig-xy}
\end{equation}%
To better understand the above result we should recall that, as shown in
Ref. \cite{Bernardes}, the corresponding expectation values of $x_{H}(t)$
and $y_{H}(t)$ are
\begin{eqnarray}
\langle x_{H}(t)\rangle  &=&\frac{\eta ^{2}k_{0y}}{(\hbar \Omega )^{2}}\sin
^{2}(\Omega t)\,,  \label{xh} \\
\langle y_{H}(t)\rangle  &=&\frac{\hbar k_{0y}}{m^{*}}t+\frac{\eta
^{2}k_{0y}\epsilon _{-}}{2(\hbar \Omega )^{3}}[\sin \left( 2\Omega
t\right) -2\Omega t].  \label{yh}
\end{eqnarray}
This oscillatory zitterbewegung shown in the above expectation
values is proportional to the initial group wave length, a fact
which offers way more favorable perspectives for the experimental
detection of this effect than for free or weakly bound electrons. In
the latter case the amplitude of the zitterbewegung is of order the
free-electron Compton wave length and therefore by orders of
magnitude smaller. For further details we refer to the discussion in
Ref.~\cite{zitter1}. Note also that differently from the
zitterbewegung following from the ordinary Rashba s-o coupling in a
single band \cite{zitter1}, for which the trembling motion is
perpendicular to the direction of propagation, here we find an
oscillatory contribution in $\langle y_{H}(t)\rangle$ along the
initial group velocity. From Eqs. (\ref{exp-sigz}) and (\ref{xh}) we
can see that $\langle \sigma _{H}^{z}(t)\rangle =1-2k_{0y}\langle
x_{H}(t)\rangle $. That is, the dynamics in spin and real spaces are
coupled. To make this point even more explicit, let us calculate the
time dependent expectation of the $z$ component of the orbital
angular momentum: $l_{H}^{z}(t)=U^{\dagger }l^{z}\otimes
\mathbf{1}U=U^{\dagger }(xp_{y}-yp_{x})\otimes \mathbf{1}U$. A
straightforward calculation yields
\begin{equation}
\langle l_{H}^{z}(t)\rangle =\frac{\hbar }{2}\frac{2\eta ^{2}k_{0y}^{2}}{%
\hbar ^{2}\Omega ^{2}}\sin ^{2}(\Omega t)=\hbar k_{0y}\langle
x_{H}(t)\rangle   \label{lzh}
\end{equation}%
Since $\vec{s}=\hbar \vec{\sigma}/2$ defines the spin angular momentum, we
can immediately see that
\begin{equation}
\langle s_{zH}(t)\rangle +\langle l_{zH}(t)\rangle =\hbar /2.  \label{szlz}
\end{equation}%
The above result is easily undertood when we recall that at $t=0$ we
injected a spin-up wave with \emph{zero} angular momentum (i.e., at $t=0$  $%
j_{z}=s_{z}=\hbar /2$). The conservation law in (\ref{szlz}) is
general (i.e., valid not just for the expectation values) and
follows from the fact that $[l_{z}+s_{z},\mathcal{H}]=0$.\ The
constraint $\dot{l_{z}}=-\dot{s_{z}} $ together with the s-o
coupling strongly tie the dynamics of the system thus leading to the
zitterbewegung in both the spin and real variables.

\section{Summary}

In summary, we have derived expressions for the zitterbewegung in
the spin variables due to the inter-subband-induced s-o coupling
present in symmetric wells with two subbands. For the simple case of
an injected spin-up plane wave, we explicitly show that the
zitterbewegung in both the spin and real spaces are intrinsically
linked. This follows from both the s-o coupling and the conservation
of the \textit{z} component of the total angular momentum in our
system.

\begin{acknowledgements}
The authors acknowledge useful discussions with S. Erlingsson, D. S.
Saraga, M. Lee, D. Bulaev, J. Lehmann and M. Duckheim. This work was
supported by the Swiss NSF, the NCCR Nanoscience, EU NoE MAGMANet,
DARPA, ARO, ONR, JST ICORP, CNPq, FAPESP, and DFG via SFB 689.
\end{acknowledgements}

\end{document}